\documentclass[10pt,preprintnumbers,showpacs,amsmath,amssymb,cite,onecolumn]{revtex4}
\usepackage{graphicx}
\usepackage{txfonts}
\usepackage{bm}   
\usepackage{amsmath,mathrsfs}
\usepackage{multirow}
\begin{document}

\title{A determination of the flavor asymmetric sea quarks in the proton}
\author{
Wei Zhu$^1$,
Rong Wang$^{2,3}$,
Jianhong Ruan$^1$,
Xurong Chen$^{2}$,
Pengming Zhang$^{2}$
}
\affiliation{
$^1$ Department of Physics, East China Normal University, Shanghai 200062, China\\
$^2$ Institute of Modern Physics, Chinese Academy of Sciences, Lanzhou 730000, China\\
$^3$ Lanzhou University, Lanzhou 730000, China\\
}

\begin{abstract}
Using DGLAP evolution equation with the corrections of parton recombinations,
the flavor asymmetric sea quark distributions in the proton are
first extracted from the available experimental data with the global QCD
analysis method. Comparisons of our predictions to the CTEQ, MRS, and
GRV parton distribution functions for sea quarks are presented.
Based on the separation of the flavor symmetric and asymmetric sea,
the possible relations between strange quark distribution
and symmetric light quark distribution are discussed. The results are used
to explain the recent HERMES result of the strange quark distribution.
\end{abstract}

\maketitle

\section{Introduction}
\label{SecI}

The observation of the violation of Gottfried sum rule \cite{Gott}
from the deep inelastic scattering (DIS) experiment \cite{NMC-Gott,NMC-Gott-2},
the surprisingly large asymmetry between the up (u) and down (d)
antiquark distributions in the proton measured from the
Drell-Yan pair production \cite{E866-1998,E866-2001}
and the semi-inclusive \cite{HERMES-1998} processes
revealed a new light on the origins of the nucleon sea.
While we still have not an adequate theory for the flavor asymmetry of
sea quarks, the flavor asymmetric sea as a new recognized partonic component
should be extracted from the experimental data in order to
gain deeper and more precise insights of the nucleon structure.
Although the sea quark distributions of several widely used
parton distribution functions (PDF) \cite{cteq4,MRS96,GRV98} of the proton are updated
using the Dokshitzer-Gribov-Lipatov-Altarelli-Parisi (DGLAP) equation \cite{Dokshitzer,Gribov,Altarelli}
to the experimental measurements, the symmetric and asymmetric sea quark
components have not been completely separated yet.
The solutions of QCD evolution equation depend on the initial parton
distributions at an arbitrary low scale $Q_0^2\sim 1$ GeV$^2$.
The difficulty is that the asymmetric sea always mixes
with the symmetric sea since the gluons are already produced at the scale $Q_0^2$.

The flavor asymmetric sea is created by the nonperturbative QCD
process since the perturbative processes cannot cause a significant
difference between $\overline{d}$ and $\overline{u}$.
One novel way to separate the asymmetric sea from the symmetric
sea is to perform the QCD evolution starting from an low enough resolution
scale $\mu^2$ where the nucleon contains only the nonperturbative
valence and asymmetric sea quarks while the flavor symmetric sea quarks
and gluons are not produced at this initial scale.
Once the asymmetric sea distributions at $\mu^2$ are determined,
they will be given at any $Q^2>\mu^2$, which can be used to explain
the experiments or to make predictions.

It has been known that the application of the standard DGLAP equation
is invalid at $Q^2<1$ GeV$^2$ and this equation should be modified at
small $x$ and low $Q^2$ due to the correlations among partons.
These nonlinear effects among partons to the DGLAP evolution
have been shown in the perturbative QCD \cite{GLR,MQ}.
Particularly, the present theory including the complete parton
recombinations in the leading order (LO) approximation are derived by
Zhu, Ruan and Shen (ZRS) \cite{Zhu,ZhuRuan,ZhuShen}.
Numerical calculations \cite{XChen} show that the negative nonlinear corrections
can slow down the gluon splitting and improve the perturbative stability
of the QCD evolution at low $Q^2$.
Although the contributions of the nonperturbative QCD effects
to the measured structure functions at $\mu^2<Q^2<1$ GeV$^2$
(for example, the hadronic components in the virtual photon)
are necessary, these nonperturbative effects are strongly $Q^2$-power suppressed
and they don't participate in the evolution of the parton distributions.
It is found that the predicted parton distributions at $Q^2>1$ GeV$^2$
of this approach are consistent with the experiments.

In this work, the asymmetric light sea and the valence quark distributions
at the initial scale $\mu^2$ in the proton are determined from the experimental data.
With the nonlinear corrections considered, parton distributions are evolved
starting from a low resolution scale $\mu^2$, where the nucleon consists
entirely of the nonperturbative components. All symmetric sea quarks and gluons
are radioactively produced from the QCD fluctuations.
The asymmetric sea is completely separated from the symmetric sea.
With the separated symmetric light sea distributions, we speculate
some possible relations between the strange quark distribution and the
symmetric light quark distribution. These results can be
used to check the predictions of various models for the flavor
structure of the proton and to improve the current PDFs.
The organization of the paper is as follows.
Section \ref{SecII} presents the analysis method and the ZRS corrections
to the DGLAP evolution equation. Comparisons with the experimental data
and the predictions of various models are presented in Sec. \ref{SecIII}.
Discussions and a summary are given in Sec. \ref{SecIV}.
The simplification of the complete nonlinear corrections
is presented in the appendix.

\section{Analysis method}
\label{SecII}

In this analysis, $q^v(x,Q^2)$ (q=u,d) denote the valence quark distributions,
$q^s(x,Q^2)=\overline{q}^s(x,Q^2)$ (q=u,d,s...) denote the
symmetric sea quark distributions,
$q^{as}(x,Q^2)=\overline{q}^{as}(x,Q^2)$ ( q=u,d) denote
the asymmetric sea quark distributions, and
$g(x,Q^2)$ denotes the gluon distribution.
The flavor non-singlet (NS) distribution is defined as
$q^{NS}(x,Q^2)=[q^v(x,Q^2)+q^{as}(x,Q^2)+\overline{q}^{as}(x,Q^2)]$,
and the flavor-singlet distribution is defined as
$\Sigma(x,Q^2)=\sum_q[q^{NS}(x,Q^2)+q^s(x,Q^2)+\overline{q}^{s}(x,Q^2)]$.
We take the parton-antiparton symmetry for the sea quarks
and the isospin symmetry for the flavor symmetric sea $\overline{u}^s=\overline{d}^s$.
The possible asymmetry for the strange sea $s(x)\neq\overline{s}(x)$
is not considered in this work since it refers to a different mechanism.
The present study is focused only on the flavor asymmetric light sea.

The $F_2$ \cite{H1-96-97,ZEUS-96,H1-00-01,ZEUS-01,NMC-97,E665},
$xF_3$ \cite{CCFR,CDHSW,WA59}, $\overline{d}-\overline{u}$ and $\overline{d}/\overline{u}$
\cite{E866-1998,E866-2001,HERMES-1998} data are used to extract the flavor asymmetric light sea component.
The $F_2(x,Q^2)$ data are taken to determine the non-singlet quark distributions
$u^{NS}(x,Q^2)$ and $d^{NS}(x,Q^2)$.
To separate $u^{as}(x)$ and $d^{as}(x)$ from $u^v(x)$ and $d^v(x)$,
we can use the average structure function of the neutrino and antineutrino scattering
on the iso-scalar nuclei,
\begin{equation}
\begin{aligned}
xF_3(x,Q^2)=\frac{1}{2}[xF_3^{\nu N}(x,Q^2)+xF_3^{\overline{\nu} N}(x,Q^2)]
=xu^v(x,Q^2)+xd^v(x,Q^2).
\end{aligned}
\label{xF3-def}
\end{equation}
Thus we have
\begin{equation}
\begin{aligned}
u^{as}(x,Q^2)+d^{as}(x,Q^2)=u^{NS}(x,Q^2)+d^{NS}(x,Q^2)-u^v(x,Q^2)-d^v(x,Q^2)\\
=u^{NS}(x,Q^2)+d^{NS}(x,Q^2)-xF_3(x,Q^2).
\end{aligned}
\label{uasAdddas}
\end{equation}
On the other hand,
\begin{equation}
\begin{aligned}
d^{as}(x,Q^2)-u^{as}(x,Q^2)=
\overline{d}(x,Q^2)-\overline{u}(x,Q^2),
\end{aligned}
\label{dasMinusuas}
\end{equation}
are already measured in the E866 \cite{E866-1998,E866-2001} and HERMES \cite{HERMES-1998} experiments.
In principle, the initial asymmetric light sea distributions can be directly determined
combining Eqs. (\ref{uasAdddas}) and (\ref{dasMinusuas}).
Unfortunately, the distributions of the asymmetric sea
quarks are much smaller than that of the valence quarks.
Therefore, the experimental errors of $F_2$ and $xF_3$
may hinder us to extract the distributions $q^{as}$.
Instead, the initial parton distributions in this work are parametrized
and determined by a global QCD fit to the experimental data of
$F_2$, $xF_3$, $\overline{d}-\overline{u}$ and $\overline{d}/\overline{u}$.

The parametrization of the initial parton distributions is written as
\begin{equation}
\begin{aligned}
&xu^v(x,\mu^2)=Ax^{B}(1-x)^{C}(1+D\sqrt{x}+Ex), \\
&xd^v(x,\mu^2)=F(1-x)^{G}xu^v, \\
&x\overline{u}^{as}(x,\mu^2)=
  A^{uas}x^{B^{uas}}(1-x)^{C^{uas}}[(1-x)^{D^{uas}}+E^{uas}], \\
&x\overline{d}^{as}(x,\mu^2)=
  A^{das}x^{B^{das}}(1-x)^{C^{das}}[(1-x)^{D^{das}}+E^{das}].
\end{aligned}
\label{input-for}
\end{equation}
At the initial scale $\mu^2$, there are no symmetric sea quarks and gluons,
which is written as
\begin{equation}
\begin{aligned}
&u^s(x,\mu^2)=\overline{u}^s(x,\mu^2)=0, \\
&d^s(x,\mu^2)=\overline{d}^s(x,\mu^2)=0, \\
&s(x,\mu^2)=\overline{s}(x,\mu^2)=0, \\
&g(x,\mu^2)=0.
\end{aligned}
\label{input-symmSea-gluon}
\end{equation}
Besides, the initial parton distributions are constrained
by the two following sum rules, which are expressed in the form
\begin{equation}
\int_0^1 dxx[u^{NS}(x,\mu^2)+d^{NS}(x,\mu^2)]=1,
\label{momSR}
\end{equation}
for the momenta of partons and
\begin{equation}
\begin{aligned}
&\int_0^1dxu^{NS}(x,Q^2)=2+2<u^{as}>_1, \\
&\int_0^1dxd^{NS}(x,Q^2)=1+2<d^{as}>_1,
\end{aligned}
\label{NumSR}
\end{equation}
for the initial numbers of the nonperturbative partons,
where $<q>_1$ is the first moment of the distribution $q$.
In the global fit, the value $<d^{as}>_1-<u^{as}>_1$ is fixed at 0.118
by the experimental measurement \cite{E866-2001}.

DGLAP equation with ZRS corrections at leading order \cite{Zhu,ZhuRuan,ZhuShen}
is applied in the work to evolve the nonperturbative initial parton distributions.
Here we use the simplified form of the equation
for simplicity (see appendix), which is expressed as
\begin{equation}
\begin{aligned}
&Q^2\frac{dxq^{NS}(x,Q^2)}{dQ^2} \\
&=\frac{\alpha_s(Q^2)}{2\pi}P_{qq}\otimes q^{NS},
\end{aligned}
\label{ZRS-NS}
\end{equation}
for the flavor non-singlet quarks,
\begin{equation}
\begin{aligned}
&Q^2\frac{dxq^s(x,Q^2)}{dQ^2} \\
&=\frac{\alpha_s(Q^2)}{2\pi}[P_{qq}\otimes q^s+P_{qg}\otimes g] \\
&-\frac{\alpha_s^2(Q^2)}{4\pi R^2 Q^2}\int_x^{1/2}\frac{dy}{y}x
  P_{gg\rightarrow q}(x,y)[ yg(y,Q^2)]^2 \\
&+\frac{\alpha_s^2(Q^2)}{4\pi R^2 Q^2}\int_{x/2}^x\frac{dy}{y}x
  P_{gg\rightarrow q}(x,y)[ yg(y,Q^2)]^2, (if~x\le 1/2), \\
&Q^2\frac{dxq^s(x,Q^2)}{dQ^2}  \\
&=\frac{\alpha_s(Q^2)}{2\pi}[P_{qq}\otimes q^s+P_{qg}\otimes g] \\
&+\frac{\alpha_s^2(Q^2)}{4\pi R^2 Q^2}\int_{x/2}^{1/2}\frac{dy}{y}x
P_{gg\rightarrow q}(x,y)[ yg(y,Q^2)]^2,(if~1/2\le x\le1),
\end{aligned}
\label{ZRS-sea}
\end{equation}
for the symmetric sea quarks, and
\begin{equation}
\begin{aligned}
&Q^2\frac{dxg(x,Q^2)}{dQ^2} \\
&=\frac{\alpha_s(Q^2)}{2\pi}[P_{gq}\otimes \Sigma+P_{gg}\otimes g] \\
&-\frac{\alpha_s^2(Q^2)}{4\pi R^2 Q^2}\int_x^{1/2}\frac{dy}{y}x
  P_{gg\rightarrow g}(x,y)[ yg(y,Q^2)]^2 \\
&+\frac{\alpha_s^2(Q^2)}{4\pi R^2 Q^2}\int_{x/2}^x\frac{dy}{y}x
  P_{gg\rightarrow g}(x,y)[yg(y,Q^2)]^2,(if~x\le 1/2), \\
&Q^2\frac{dxg(x,Q^2)}{dQ^2} \\
&=\frac{\alpha_s(Q^2)}{2\pi}[P_{gq}\otimes \Sigma+P_{gg}\otimes g] \\
&+\frac{\alpha_s^2(Q^2)}{4\pi R^2 Q^2}\int_{x/2}^{1/2}\frac{dy}{y}x
  P_{gg\rightarrow g}(x,y)[ yg(y,Q^2)]^2,(if~1/2\le x\le1),
\end{aligned}
\label{ZRS-gluon}
\end{equation}
for the gluon, where the factor $1/(4\pi R^2)$ is from the
normalizing of the two-parton distribution, and R is the correlation length
of the two initial partons. The linear evolution terms are from standard DGLAP evolution
and the recombination functions of the nonlinear terms are expressed as
\begin{equation}
\begin{aligned}
&P_{gg\rightarrow g}(x,y)=
  \frac{9}{64}\frac{(2y-x)(72y^4-48xy^3+140x^2y^2-116x^3y+29x^4)}{xy^5}, \\
&P_{gg\rightarrow q}(x,y)=P_{gg\rightarrow \overline{q}}(x,y)=
  \frac{1}{96}\frac{(2y-x)^2(18y^2-21xy+14x^2)}{y^5}.
\end{aligned}
\label{nonlinear-term}
\end{equation}
Although a next-to-leading order (NLO) evolution is preferred,
the LO evolution is the important first step for the determination
of the initial parton distributions.

\section{Results}
\label{SecIII}

\begin{figure}[htp]
\begin{center}
\includegraphics[width=0.45\textwidth]{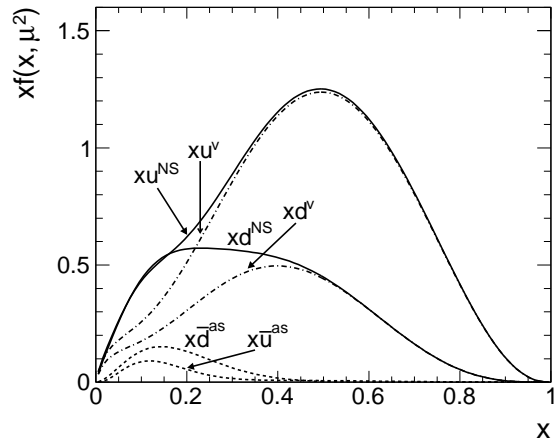}
\caption{
The obtained initial parton distributions at $\mu^2=0.064$ GeV$^2$.
}
\label{input}
\end{center}
\end{figure}

By the global fit to the experimental data,
the initial nonperturbative parton distributions are obtained to be
\begin{equation}
\begin{aligned}
&xu^v(x,\mu^2)=9.09x^{0.944}(1-x)^{2.38}(1-4.3\sqrt{x}+6.8x), \\
&xd^v(x,\mu^2)=0.765(1-x)^{1.1}xu_v, \\
&x\overline{u}^{as}(x,\mu^2)=294x^{2.62}(1-x)^{3.4}[(1-x)^{16.5}+0.00147], \\
&x\overline{d}^{as}(x,\mu^2)=42.5x^{1.97}(1-x)^{7.5}[(1-x)^{4.2}+10^{-5}], \\
\end{aligned}
\label{obtained-input}
\end{equation}
which are shown in Fig. \ref{input}. The first moment of asymmetric
down quark is $<d^{as}>_1=0.261$ with $<d^{as}>_1-<u^{as}>_1=0.118$.
The other two free parameters in Eqs.
(\ref{ZRS-NS}-\ref{ZRS-gluon}) are determined to be
$R=3.977$ GeV$^{-1}$, $\mu^2=0.064$ GeV$^2$ with $\Lambda_{QCD}=0.204$ GeV.

\begin{figure}[htp]
\begin{center}
\includegraphics[width=0.6\textwidth]{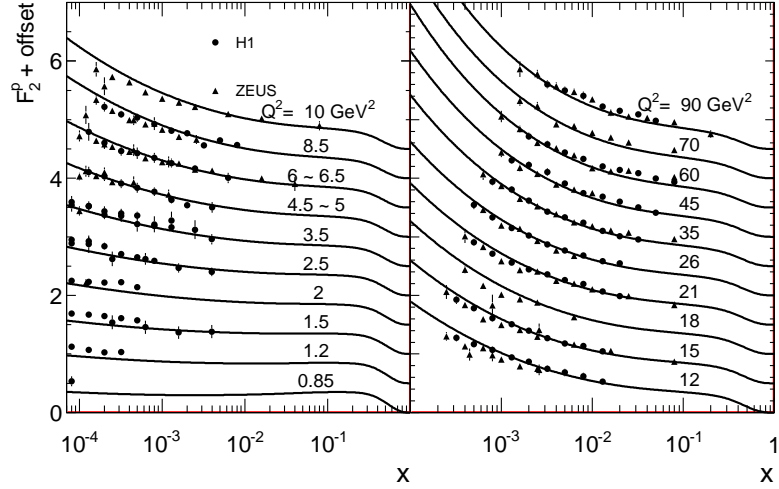}
\caption{
Comparisons of the predicted $x$-dependence of $F_2$ to the experimental data
from H1 \cite{H1-96-97} and ZEUS \cite{ZEUS-96} at high $Q^2$.
}
\label{x-dependence}
\end{center}
\end{figure}

\begin{figure}[htp]
\begin{center}
\includegraphics[width=0.6\textwidth]{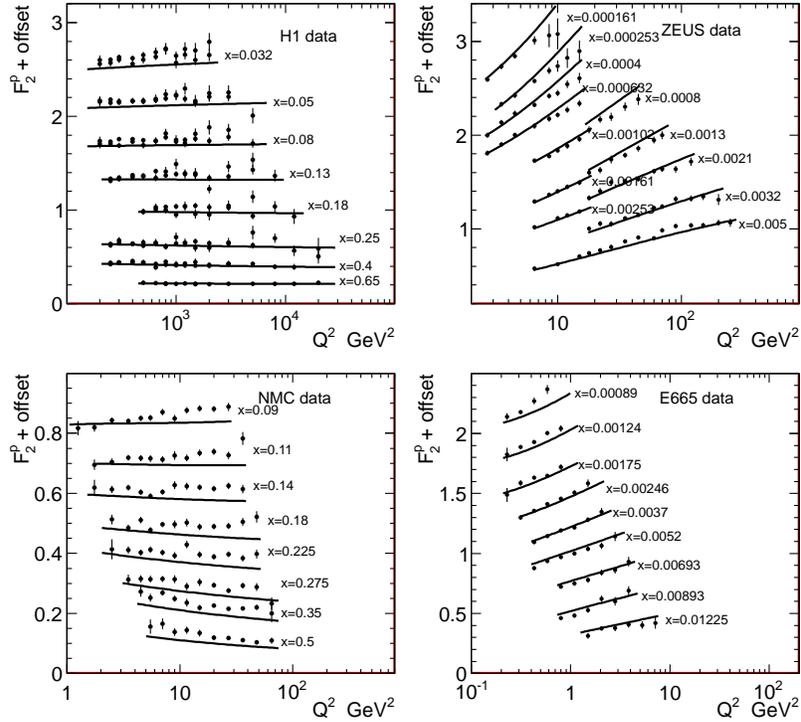}
\caption{
Comparisons of the predicted $Q^2$-dependence of $F_2$ to the experimental data
from H1 \cite{H1-00-01}, ZEUS \cite{ZEUS-01},
NMC \cite{NMC-97} and E665 \cite{E665} at large and small $x$.
}
\label{q2-dependence}
\end{center}
\end{figure}

\begin{figure}[htp]
\begin{center}
\includegraphics[width=0.45\textwidth]{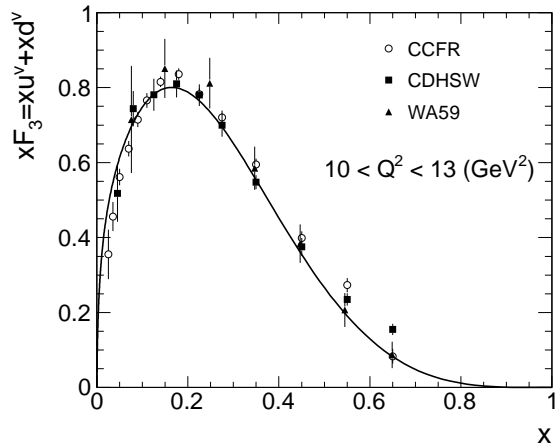}
\caption{
Comparisons of the predicted $xF_3$ to the experimental data
from CCFR \cite{CCFR}, CDHSW \cite{CDHSW} and WA59 \cite{WA59}.
}
\label{xF3}
\end{center}
\end{figure}

\begin{figure}[htp]
\begin{center}
\includegraphics[width=0.45\textwidth]{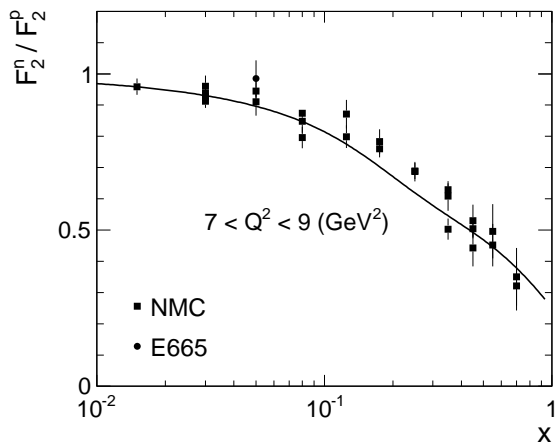}
\caption{
Comparisons of the predicted structure function ratios
$F_2^n/F_2^p$ to the experimental data
from NMC \cite{NMC-90} and E665 \cite{E665-95}.
}
\label{F2n-over-F2p}
\end{center}
\end{figure}

\begin{figure}[htp]
\begin{center}
\includegraphics[width=0.45\textwidth]{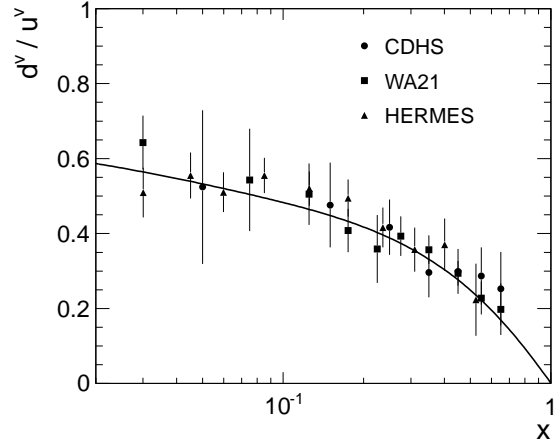}
\caption{
Comparisons of the predicted valence quark ratios
$d^v/u^v$ to the measured data ($Q^2=2.4-42.9$ GeV$^2$)
from CDHS \cite{CDHS}, WA21 \cite{WA21} and HERMES \cite{HERMES}.
}
\label{dv-over-uv}
\end{center}
\end{figure}

\begin{figure}[htp]
\begin{center}
\includegraphics[width=0.45\textwidth]{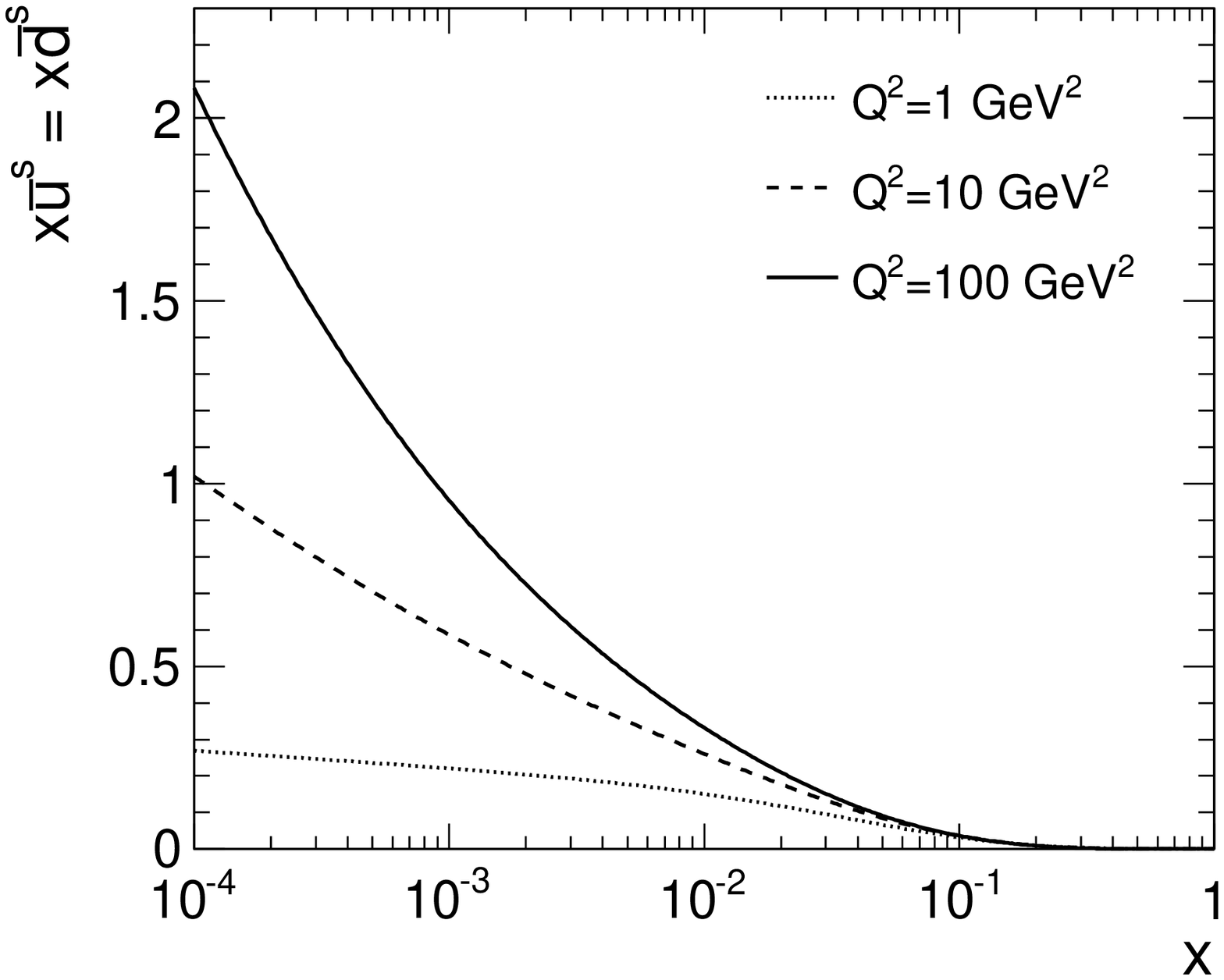}
\caption{
The predicted symmetric light sea quark distributions in the proton
at different resolution scales.
}
\label{SymmSea}
\end{center}
\end{figure}

\begin{figure}[htp]
\begin{center}
\includegraphics[width=0.45\textwidth]{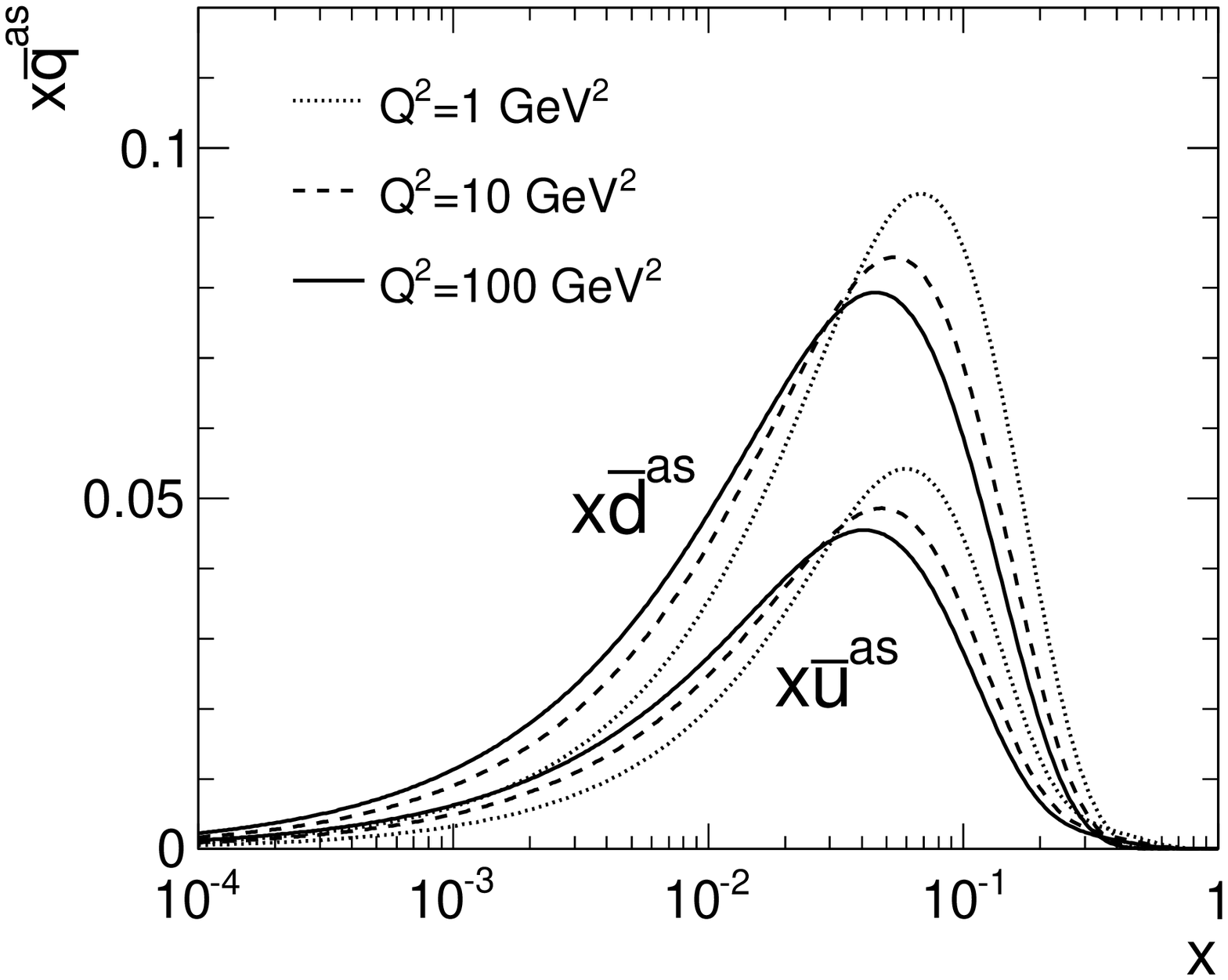}
\caption{
The predicted asymmetric light sea quark distributions in the proton
at different resolution scales.
}
\label{AsymmSea}
\end{center}
\end{figure}

\begin{figure}[htp]
\begin{center}
\includegraphics[width=0.5\textwidth]{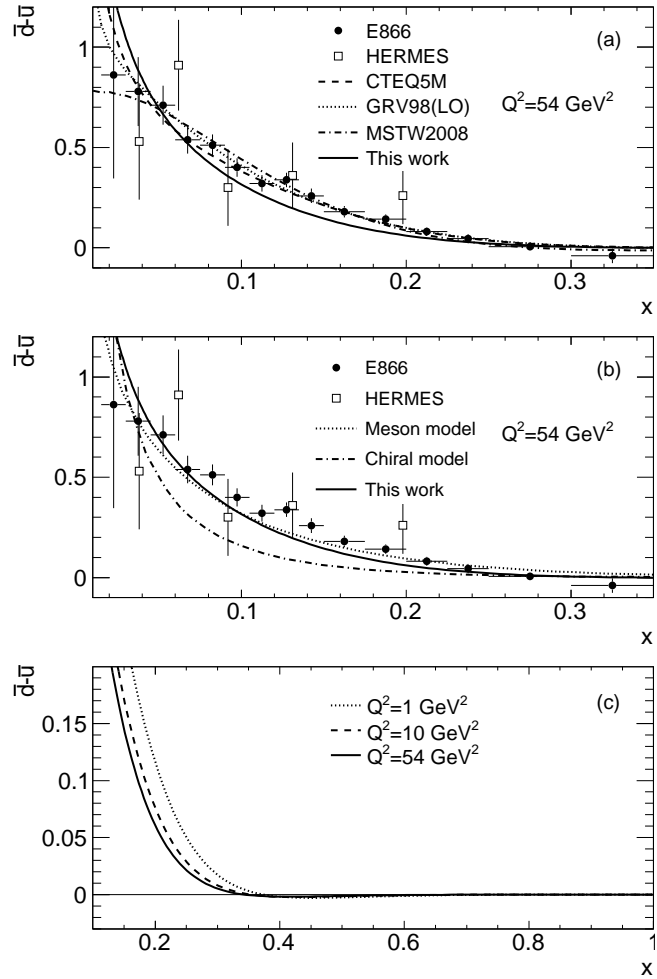}
\caption{
Comparisons of the predicted light sea quark differences
$\overline{d}-\overline{u}$ in the proton as a function of $x$ at
$Q^2=54$ GeV$^2$ to the experimental data from
Fermilab E866 \cite{E866-1998,E866-2001} and HERMES \cite{HERMES-1998}.
(a) comparisons with the popular PDFs from CTEQ5M \cite{cteq5},
GRV98 \cite{GRV98} and MSTW \cite{mstw08};
(b) comparisons with meson and chiral models \cite{Melnitchouk};
(c) the possible sign-change of $\overline{d}(x)-\overline{d}(x)$
around $x=0.3$ is shown.
}
\label{dbar-ubar}
\end{center}
\end{figure}

\begin{figure}[htp]
\begin{center}
\includegraphics[width=0.5\textwidth]{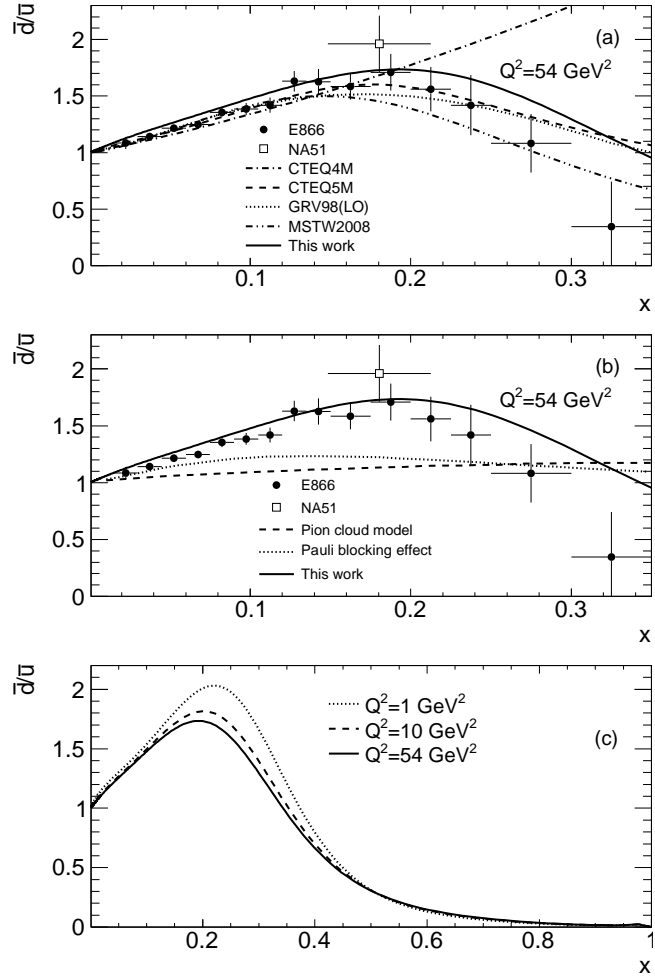}
\caption{
Comparisons of the predicted light sea quark ratios
$\overline{d}/\overline{u}$ in the proton as a function of $x$ at
$Q^2=54$ GeV$^2$ to the experimental data from
Fermilab E866\cite{E866-1998,E866-2001} and NA51 \cite{NA51}.
(a) comparisons with the popular PDFs from CTEQ4M \cite{cteq4},
CETQ5M \cite{cteq5}, GRV98 \cite{GRV98} and MSTW \cite{mstw08};
(b) comparisons with meson model \cite{Melnitchouk};
(c) the ratios $\overline{d}/\overline{u}$ at large $x$.
}
\label{dbar-over-ubar}
\end{center}
\end{figure}

\begin{figure}[htp]
\begin{center}
\includegraphics[width=0.45\textwidth]{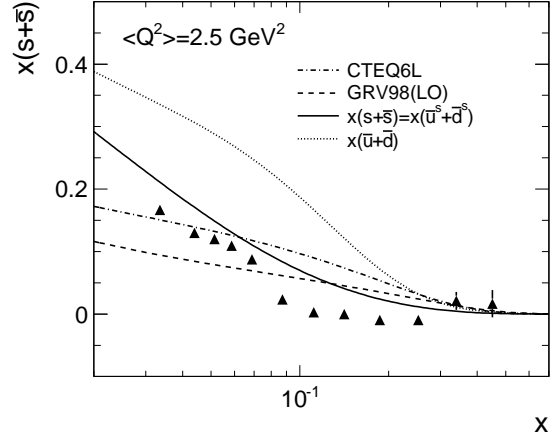}
\caption{
Our predicted strange quark distribution
$x[s(x,Q^2)+\overline{s}(x,Q^2)]$ at $Q^2=2.5$ GeV$^2$
compared to the HERMES data \cite{HERMES-2014} and the global analyses from
CTEQ6L \cite{cteq6} and GRV \cite{GRV98}.
}
\label{s-sea}
\end{center}
\end{figure}

\begin{figure}[htp]
\begin{center}
\includegraphics[width=0.45\textwidth]{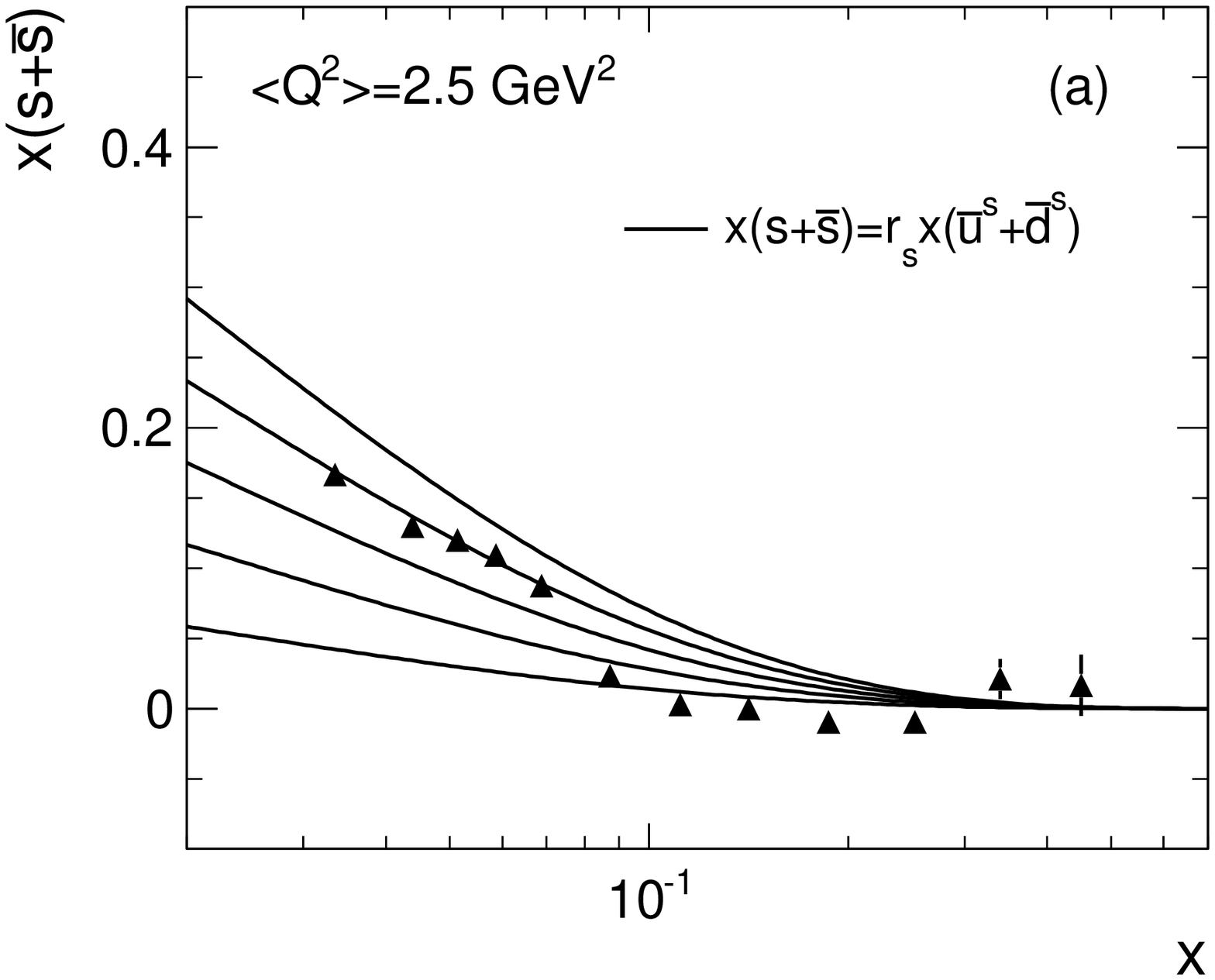}
\includegraphics[width=0.45\textwidth]{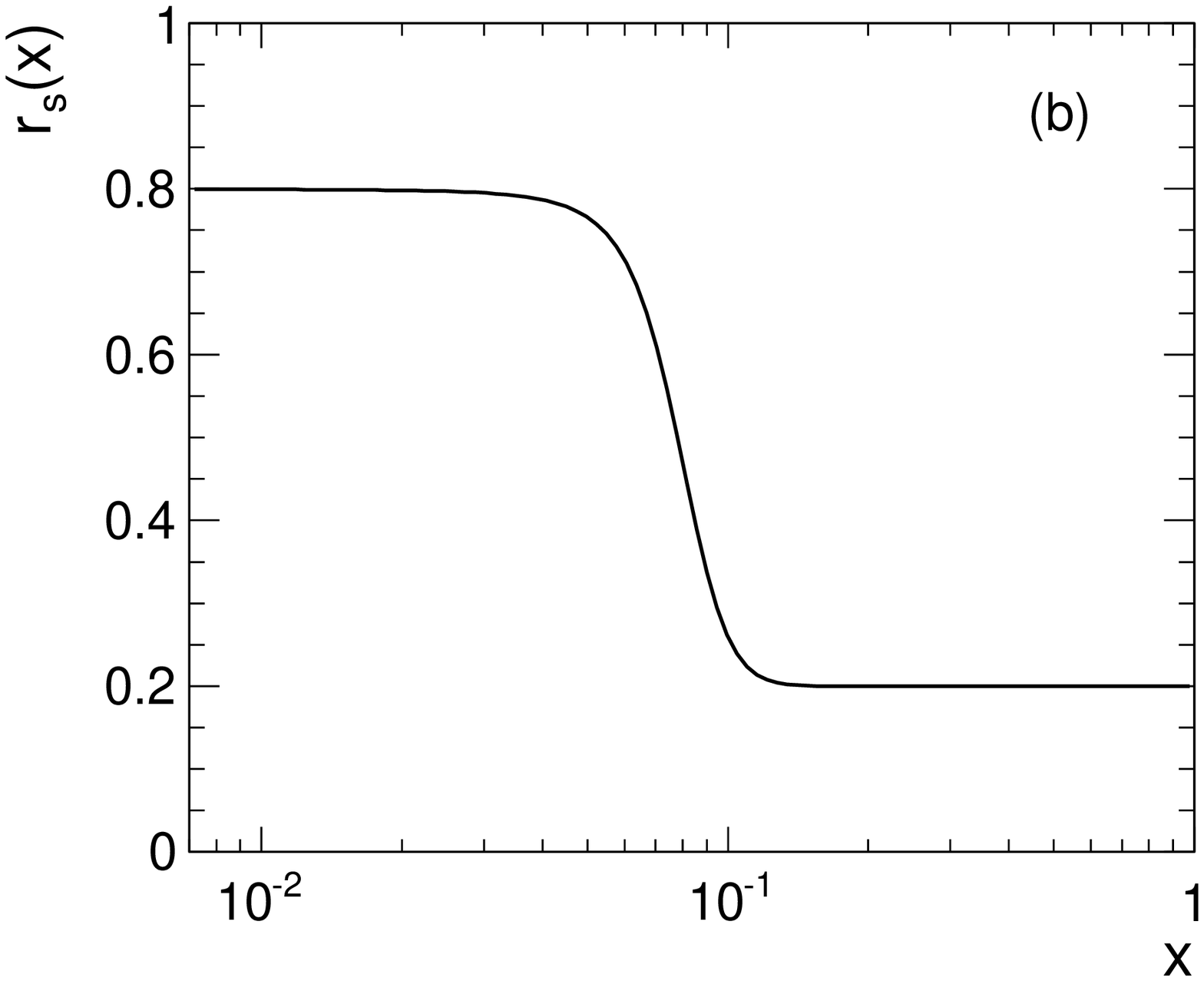}
\caption{
(a) the strange quark distribution $x[s(x,Q^2)+\overline{s}(x,Q^2)]$ at $Q^2=2.5$ GeV$^2$
with different suppression values of $r_s=1,0.8,0.6,0.4$ and $0.2$ (from top to bottom);
(b) a possible curve of the suppression ratio $r_s(x,Q^2)$ as a function of $x$
at $Q^2=2.5$ GeV$^2$.
}
\label{suppre-ratio}
\end{center}
\end{figure}

\begin{figure}[htp]
\begin{center}
\includegraphics[width=0.45\textwidth]{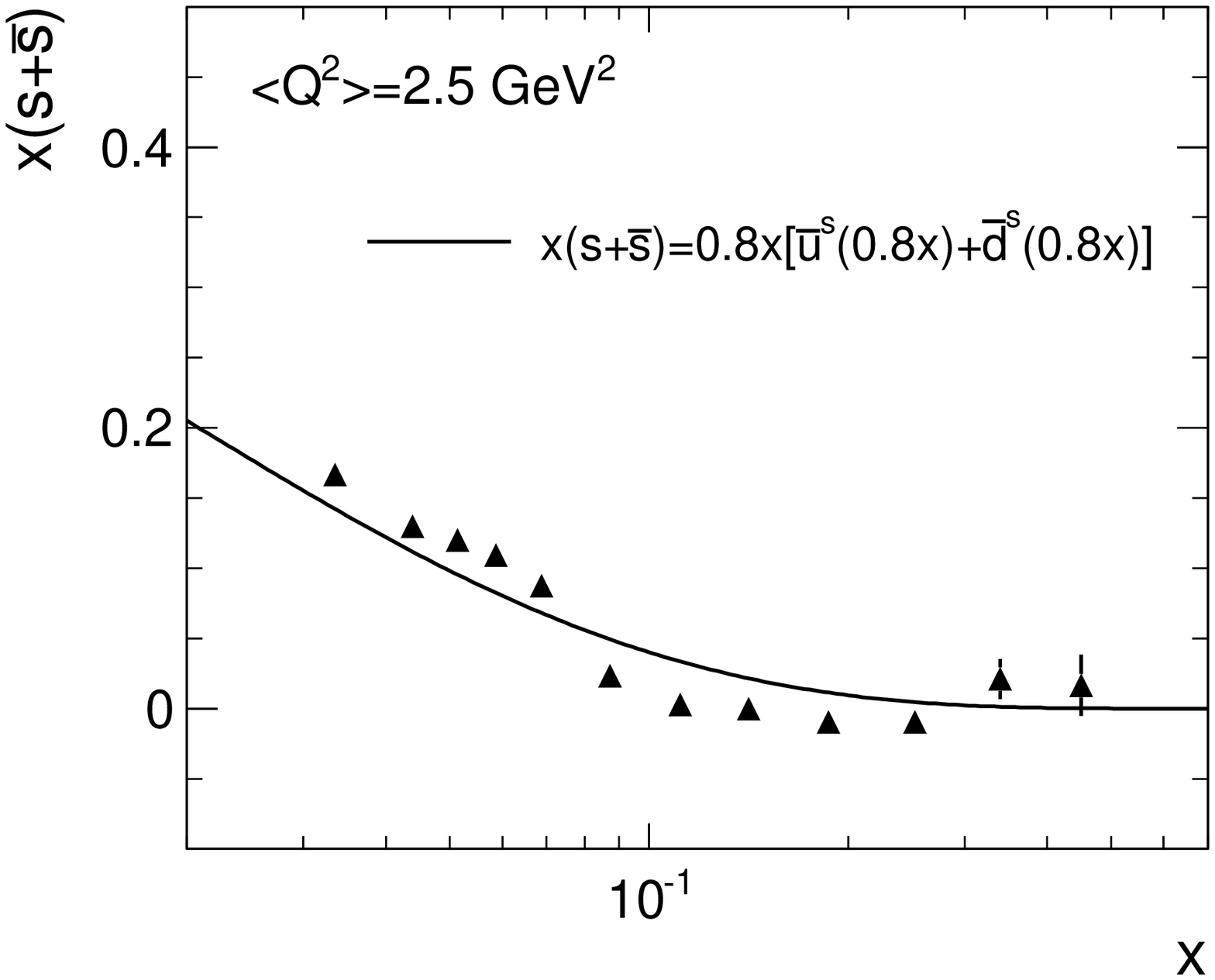}
\caption{
The strange quark distribution $x[s(x,Q^2)+\overline{s}(x,Q^2)]$ at $Q^2=2.5$ GeV$^2$
by the $x$-rescaling model.
}
\label{x-rescaling}
\end{center}
\end{figure}

\begin{figure}[htp]
\begin{center}
\includegraphics[width=0.45\textwidth]{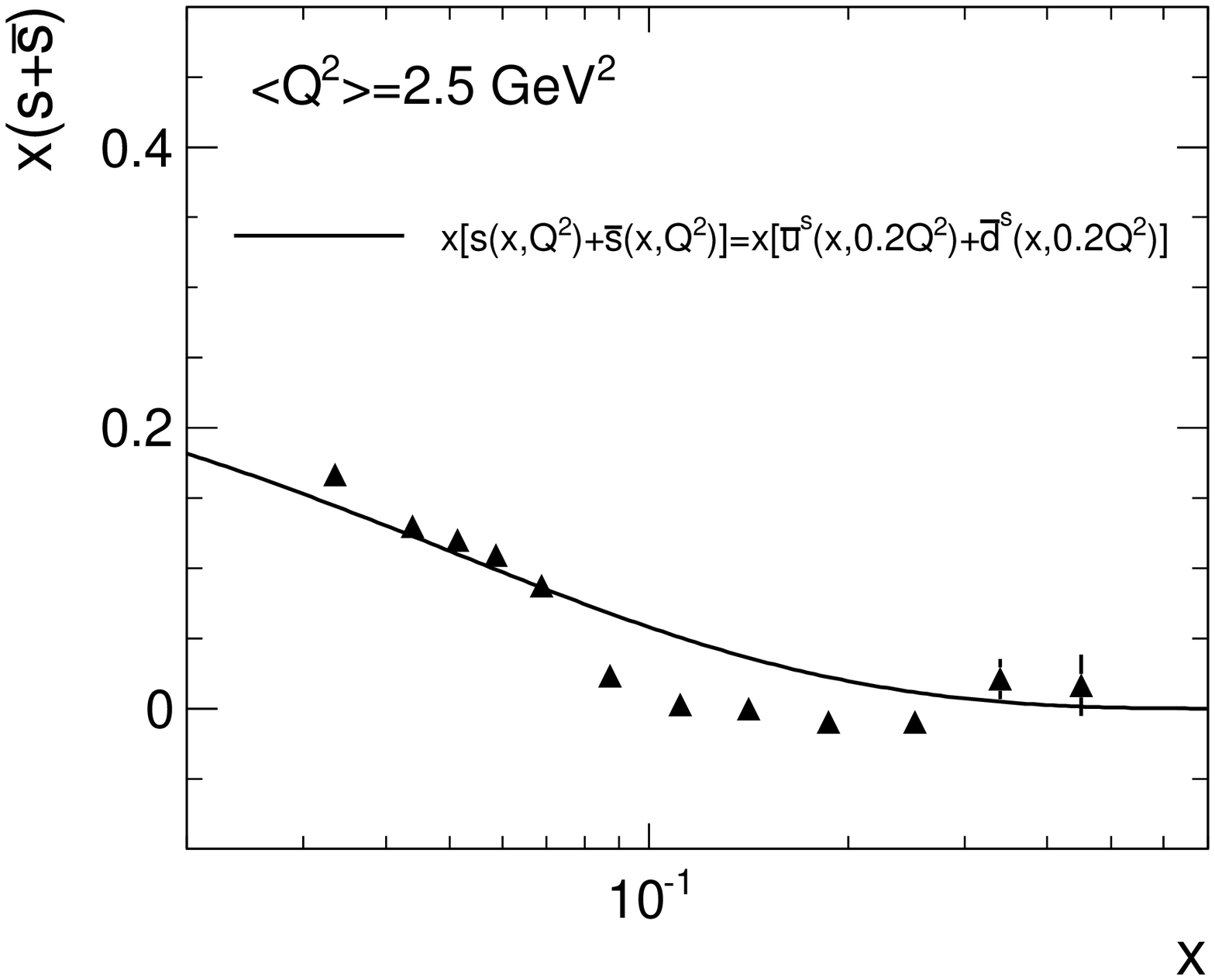}
\caption{
The strange quark distribution $x[s(x,Q^2)+\overline{s}(x,Q^2)]$ at $Q^2=2.5$ GeV$^2$
by the $Q^2$-rescaling model.
}
\label{q2-rescaling}
\end{center}
\end{figure}

\begin{figure}[htp]
\begin{center}
\includegraphics[width=0.45\textwidth]{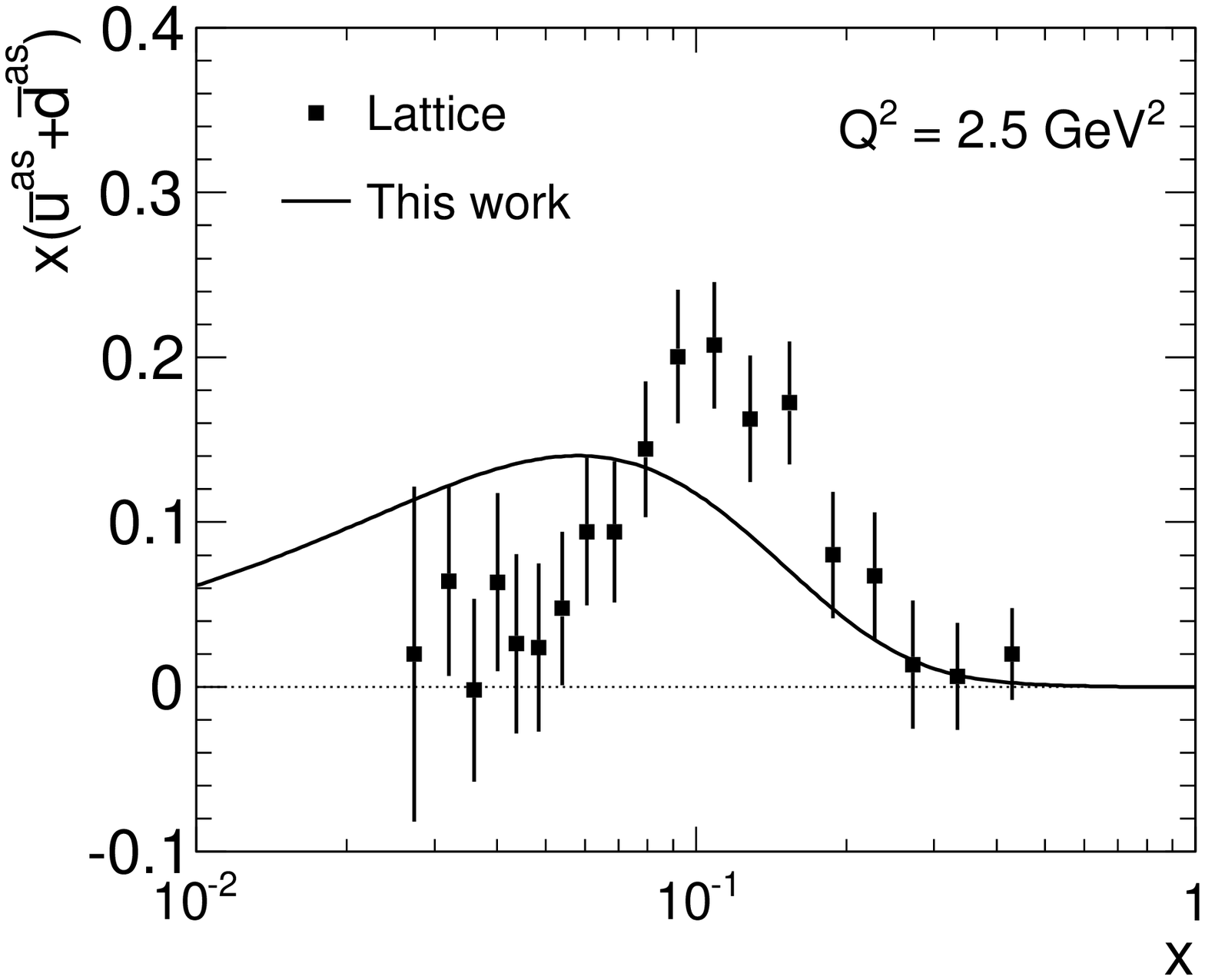}
\caption{
Comparison of our predicted asymmetric light sea quark distribution
$x[\overline{u}^{as}(x,Q^2)+\overline{d}^{as}(x,Q^2)]$ to the
connected sea quark distribution
$x[\overline{u}^{cs}(x,Q^2)+\overline{d}^{cs}(x,Q^2)]$
predicted by the lattice calculation \cite{K.F.Liu-12} at $Q^2=2.5$ GeV$^2$.
}
\label{asymm-cs}
\end{center}
\end{figure}

The $x$- and $Q^2$-dependence of the structure function
$F_2^p(x,Q^2)=x\sum_ie^2_iq_i^{NS}(x,Q^2)+x\sum_je^2_j[q_j^{s}(x,Q^2)+
\overline{q}_j^{s}(x,Q^2)]$ are shown in Figs. \ref{x-dependence} and \ref{q2-dependence},
respectively. The isoscalar
structure functions $xF_3$ provide valuable informations of
the valence quark distributions. Fig. \ref{xF3} shows the predicted $xF_3$ with the
experimental data. The ratios of the structure functions $F_2(x,Q^2)$ of
the neutron to that of the proton are sensitive to up and down quark distributions.
The calculated structure function ratios $F^n_2/F^p_2$ under the assumption of
the isospin symmetry between the proton and the neutron are shown in Fig. \ref{F2n-over-F2p}
with the experimental measurements. The predicted down to up valence quark
ratios $d^v/u^v$ are compared to the experimental data in Fig. \ref{dv-over-uv}.
Basically, our predicted structure functions and valence quark distributions
are consistent with the experimental measurements at high $Q^2$.

The obtained symmetric and asymmetric light sea quark distributions
at different $Q^2$ are shown in Figs. \ref{SymmSea} and \ref{AsymmSea}, respectively.
The predicted $\overline{d}-\overline{u}=\overline{d}^{as}-\overline{u}^{as}$
are shown in Fig. \ref{dbar-ubar}, compared to the E866 \cite{E866-1998,E866-2001} and HERMES \cite{HERMES-1998} data,
the CTEQ5M \cite{cteq5}, GRV98 \cite{GRV98} and MSTW \cite{mstw08} parton distribution functions (Fig. \ref{dbar-ubar}(a))
and some models (Fig. \ref{dbar-ubar}(b)).
The possible sign-change of $\overline{d}(x)-\overline{u}(x)$ around $x=0.3$
is an interesting and important phenomenon for understanding
the flavor structure of the nucleon sea \cite{{J.C.Peng-14}}.
Fig. \ref{dbar-ubar}(c) shows a very small negative value of $\overline{d}-\overline{u}$ in the
range about $x>0.3$, and the position where $\overline{d}-\overline{u}=0$ is $Q^2$-dependent.
The predicted ratios $\overline{d}/\overline{u}$ of the proton
are shown in Fig. \ref{dbar-over-ubar}, compared to the E866 \cite{E866-1998,E866-2001} and NA51 \cite{NA51} data,
the global fits of CTEQ4M \cite{cteq4}, CETQ5M \cite{cteq5}, GRV98 \cite{GRV98} and MSTW \cite{mstw08} (Fig. \ref{dbar-over-ubar}(a))
and some models (Fig. \ref{dbar-over-ubar}(b)).
Basically, the extracted symmetric and asymmetric light sea quark distributions
are in agreement with the experimental measurements.

Strange quark describes some important features of the sea structure of the nucleon.
However the relation between the strange sea and non-strange sea is rather
confusing \cite{Stolarski}. The decomposition of the asymmetric and symmetric sea in this work
may provide a way to study the origin of the strange sea in the proton.
As shown in Figs. \ref{SymmSea} and \ref{AsymmSea}, the asymmetric and symmetric sea
have very different $x$- and $Q^2$-dependence.
Assuming the $SU(3)_F$ flavor symmetry, we have
$s+\overline{s}=\overline{u}^s+\overline{d}^s$.
In this work, $s=\overline{s}$ is purely generated from the QCD evolution
without any initial nonperturbative components, which is shown in Fig. \ref{s-sea}
compared to the experiment. The experimental data is taken from the recent
re-evaluation of the strange quark distribution of HERMES experiment \cite{HERMES-2014}.
Although the shape of our predicted strange quark distribution is
similar to the HERMES data, the distribution is higher than the measurement.
The breaking of the $SU(3)_F$ symmetry due to a large mass of the strange
quark need to be considered.
The popular PDFs from the CTEQ6L and GRV global fits are also shown in Fig. \ref{s-sea}.
For CTEQ6L strange quark distribution \cite{cteq6}, the relation
$s(x,Q^2_0)+\overline{s}(x,Q^2_0)=0.4[\overline{u}(x,Q^2_0)+\overline{d}(x,Q^2_0)]$
is assumed at initial scale $Q^2_0=1.69$ GeV$^2$.
The strange quark distribution in the GRV analysis \cite{GRV98} is also purely dynamically
generated but starting from a high initial scale $Q_0^2=0.26$ GeV$^2$ $\gg 0.064$ GeV$^2$.
The strange quark distribution in the GRV analysis is lower than the measurement
at small $x$.

Usually the suppression ratio for strange quark
$r_s=[s(x,Q^2)+\overline{s}(x,Q^2)]/2\overline{d}(x,Q^2)\neq 1$
is used to describe the $SU(3)_F$ symmetry breaking.
However, the ratio $\overline{d}(x)/\overline{u}(x)$ is strongly dependent on $x$
due to the flavor-dependent asymmetric sea quark distributions
$\overline{d}^{as}(x)\neq\overline{u}(x)^{as}$. We should exclude
the contributions of $\overline{u}^{as}(x)$ and $\overline{d}(x)^{as}$
in the definition $r_s$. For this sake, we re-define the
suppression ratio as
\begin{equation}
\begin{aligned}
r_s(x,Q^2)=\frac{s(x,Q^2)+\overline{s}(x,Q^2)}{\overline{u}^s(x,Q^2)+\overline{d}^s(x,Q^2)}.
\end{aligned}
\label{rs-def}
\end{equation}
Strange quark distributions with different suppression values at
$Q^2=2.5$ GeV$^2$ are shown in Fig. \ref{suppre-ratio}(a).
It is found that the strange quark distribution with $r_s=0.8$
is in excellent agreement with the HERMES measurement in $x<0.07$ range.
The preferred suppression ratio $r_s=0.8$ is consistent with the ATLAS measurement \cite{ATLAS}
which obtains $r_s=1.0\pm 0.25$ at $x = 0.023$ and
the lattice QCD calculation $r_s=0.857\pm 0.040$ at $Q^2=2.5$ GeV$^2$ \cite{K.F.Liu-12}.
However, the predicted strange quark distribution at large $x>0.1$ is obviously
higher than the experimental measurement.
The suppression ratio is reduced to about 0.2 in large $x>0.1$ region
to meet the experiment.
This may suggest an unknown mechanism for the strange quark production in the nucleon.
Here, we list some phenomenological explanations: (1) the
$x$-dependent suppression ratio $r_s(x)$, which is presented in Fig. \ref{suppre-ratio}(b);
(2) the $x$-rescaling, i.e.,
$x[s(x,Q^2)+\overline{s}(x,Q^2)]=\eta x [\overline{u}^s(\eta
x,Q^2)+\overline{d}^s(\eta x,Q^2)]$ at $Q^2=2.5$ GeV$^2$ with $\eta=0.8$,
which is shown in Fig. \ref{x-rescaling};
(3) the $Q^2$-rescaling, i.e.,
$x[s(x,Q^2)+\overline{s}(x,Q^2)]=x[\overline{u}^s(x,\zeta Q^2)
+\overline{d}^s(x,\zeta Q^2)]$ with $\zeta=0.2$ at $Q^2=2.5$ GeV$^2$,
which is shown in Fig. \ref{q2-rescaling}.

In lattice QCD, the asymmetric light sea is related to the connected sea
$u^{cs}(x)$ and $d^{cs}(x)$. The connected sea component has been
determined by the calculation of lattice QCD, the previous HERMES result \cite{HERMES-08}
and the CT10 parton distribution functions \cite{CT10}, which is given by
\begin{equation}
\begin{aligned}
&\overline{u}^{cs}(x,Q^2)+\overline{d}^{cs}(x,Q^2)=[\overline{u}(x,Q^2)+\overline{d}(x,Q^2)]_{CT10}
-\frac{1}{r_s^{lattice}}[s(x,Q^2)+\overline{s}(x,Q^2)]_{pre-HERMES} \\
&=\overline{u}^{cs}(x,Q^2)+\overline{d}^{cs}(x,Q^2)+\overline{u}^{ds}(x,Q^2)+\overline{d}^{ds}(x,Q^2)
-\frac{1}{r_s^{lattice}}[s(x,Q^2)+\overline{s}(x,Q^2)]_{pre-HERMES},
\end{aligned}
\label{cs-deter}
\end{equation}
where the disconnected sea $\overline{u}^{ds}(x)$ and $\overline{d}^{ds}(x)$
are identical to our symmetric sea $\overline{u}^s(x)$ and $\overline{d}^s(x)$.
Fig. \ref{asymm-cs} shows the comparison of our obtained asymmetric sea
$x[\overline{u}^{as}(x,Q^2)+\overline{d}^{as}(x,Q^2)]$ with
the connected sea $x[\overline{u}^{cs}(x,Q^2)+\overline{d}^{cs}(x,Q^2)]$ at
$Q^2=2.5$ GeV$^2$. There are clear differences between the extracted asymmetric sea
and the connected sea. The connected sea distribution is lower than the asymmetric sea
distribution in the range about $x<0.07$, while it is higher than the asymmetric sea
distribution in the range about $x>0.07$.
The inconsistence is explained as follows.
The strange quark distribution $s(x)+\overline{s}(x)$ for the determination of
the connected sea is from the previous HERMERS data \cite{HERMES-08} which is higher than
that from the current HERMES data \cite{HERMES-2014} in the range of $x<0.07$.
As extracted by Eq. (\ref{cs-deter}), the higher strange quark distribution
results in the lower connected sea distribution in the small x region.
In the lattice calculation \cite{K.F.Liu-12}, the disconnected sea is assumed to be
$\overline{u}^{ds}(x)+\overline{d}^{ds}(x)=[s(x)+\overline{s}(x)]/r_s$ with
$r_s$ regarded as a constant. The fact is that the suppression ratio goes
very small at large $x$. Hence the determined connected sea is higher
than our extracted asymmetric sea at about $x>0.07$.
Generally, the obtained asymmetric light sea in this work is reasonable.

\section{Discussions and summary}
\label{SecIV}

The DGLAP equation with the parton recombination corrections at leading order is
applied to the low $Q^2$ region. An unavoidable question is whether
we can neglect all higher order QCD corrections when $Q^2\ll 1$ GeV$^2$?
Since all order resummation of these corrections are unable at moment,
we suggest that if the leading order contributions (or including necessary
lowest order corrections)  are compatible with the experimental data,
one may conjecture (1)  these neglected higher order corrections
may cancelable to each other; (2) or they are successfully absorbed
by a finite number of free parameters such as $\mu^2$ and $R$ in this work.
Before the confirmation of these conjectures,
our initial parton distributions at $\mu^2<1$ GeV$^2$
can be regarded as an effective nonperturbative input
which describe well the data at $Q^2>1$ GeV$^2$, as well as
to predict the asymmetry light sea quark distributions in the proton.

In summary, the flavor asymmetric sea quark distributions in the proton are
first extracted from the available experimental data
by the DGLAP evolution with the parton recombination corrections.
Moreover, with the separated flavor symmetric sea, we studied the possible
relations between the strange quark distribution and the symmetric light quark distribution.
It is found that the strange sea distribution
is different from the non-strange distributions
for that the former has few asymmetric nonperturbative components.
Strange quark distributions from the $x$-dependent suppression ratio, the $x$-rescaling
and the $Q^2$-rescaling models are basically consistent with the HERMES data.
Finally, the extracted asymmetric sea component is in agreement with
the connected sea in Lattice QCD if the re-evaluated data of the HERMES
experiment is used and the $x$-dependent suppression ratio for
the strange quark is assumed.

\section*{Acknowledgments}

One of us (W.Z.) thanks K.F. Liu for the useful discussions and suggestions.

\section*{Appendix:
simplification of ZRS corrections to the DGLAP equation}

\begin{figure}[htp]
\begin{center}
\includegraphics[width=0.5\textwidth]{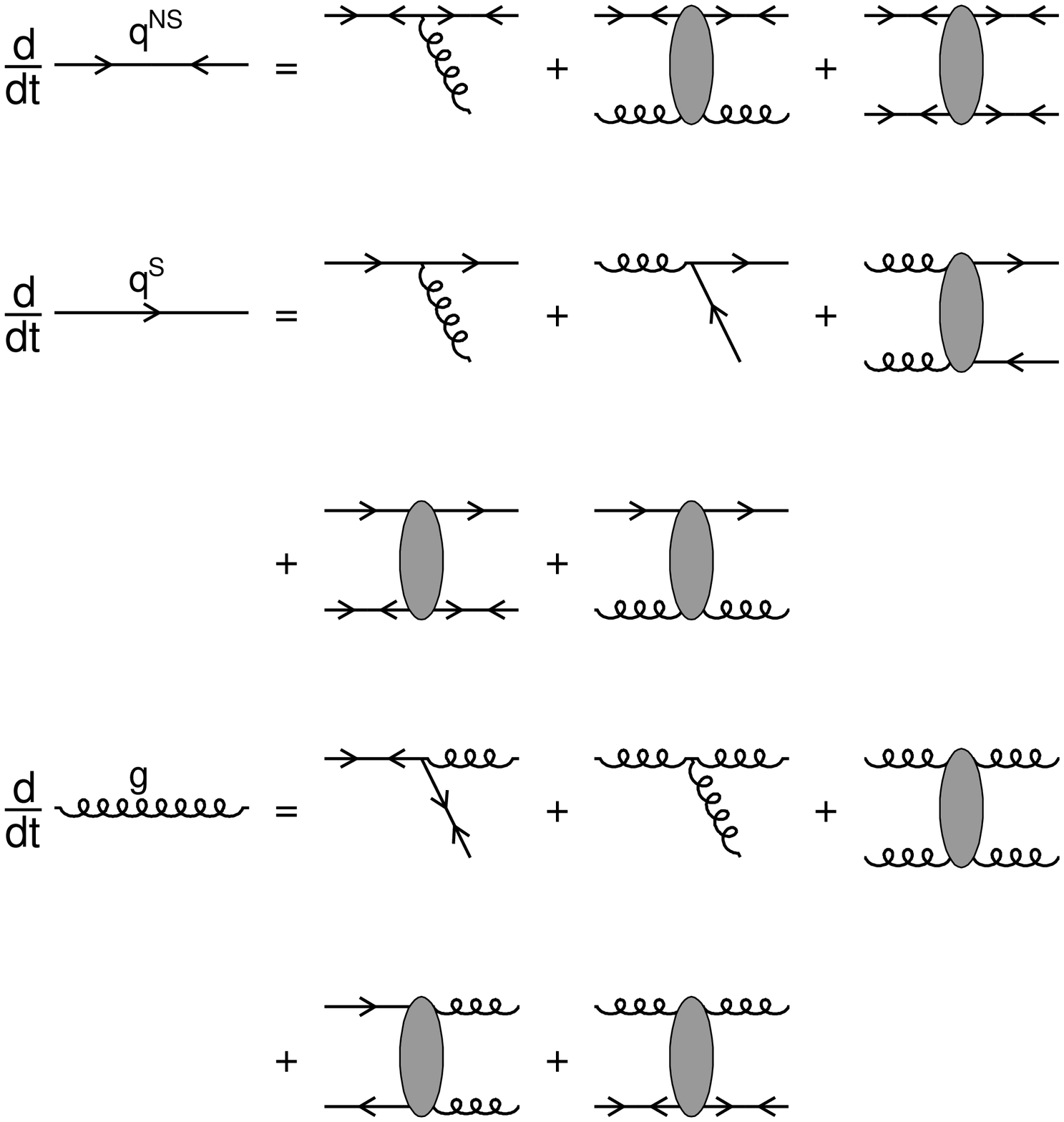}
\caption{
Diagrams of the complete parton recombination corrections to the DGLAP evolution.
}
\label{recombinations}
\end{center}
\end{figure}

\begin{figure}[htp]
\begin{center}
\includegraphics[width=0.5\textwidth]{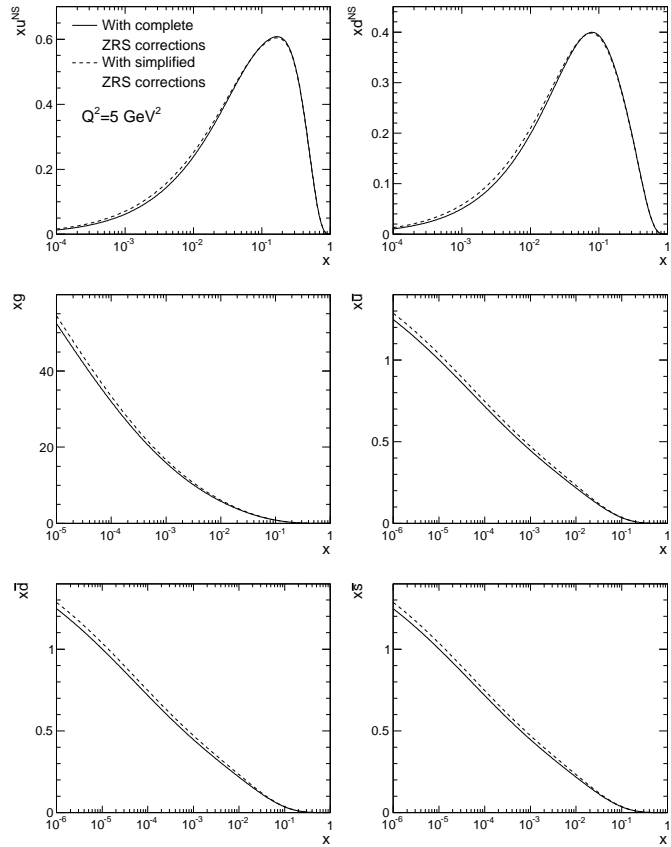}
\caption{
Comparisons between the parton distribution functions from the complete ZRS corrections
to the DGLAP evolution and that from the simplified ZRS corrections to the DGLAP evolution.
}
\label{simplification}
\end{center}
\end{figure}

The DGLAP equation with the complete ZRS
corrections at leading order \cite{Zhu,ZhuRuan,ZhuShen} according to Fig. \ref{recombinations}
are written as
\begin{equation}
\begin{aligned}
&Q^2\frac{dxq^{NS}(x,Q^2)}{dQ^2} \\
&=\frac{\alpha_s(Q^2)}{2\pi}P_{qq}\otimes q^{NS} \\
&-\frac{\alpha_s^2(Q^2)}{4\pi R^2Q^2}\int_x^{1/2}
  \frac{dy}{y}xP_{qg\rightarrow q}(x,y)yg(y,Q^2)yq^{NS}(y,Q^2) \\
&+\frac{\alpha_s^2(Q^2)}{4\pi R^2 Q^2}\int_{x/2}^x
  \frac{dy}{y}xP_{qg\rightarrow q}(x,y)yg(y,Q^2)yq^{NS}(y,Q^2) \\
&-\frac{\alpha_s^2(Q^2)}{4\pi R^2 Q^2}\int_x^{1/2}
  \frac{dy}{y}xP_{qq\rightarrow q}(x,y)y
  [\Sigma(y,Q^2)-q^{NS}(y,Q^2)]yq^{NS}(y,Q^2) \\
&+\frac{\alpha_s^2(Q^2)}{4\pi R^2 Q^2}\int_{x/2}^x
  \frac{dy}{y}xP_{qq\rightarrow q}(x,y)y
  [\Sigma(y,Q^2)-q^{NS}(y,Q^2)]yq^{NS}(y,Q^2),(if~x\le 1/2), \\
&Q^2\frac{dxq^{NS}(x,Q^2)}{dQ^2} \\
&=\frac{\alpha_s(Q^2)}{2\pi}P_{qq}\otimes q^{NS} \\
&+\frac{\alpha_s^2(Q^2)}{4\pi R^2 Q^2}\int_{x/2}^{1/2}
  \frac{dy}{y}xP_{qg\rightarrow q}(x,y)yg(y,Q^2)yq^{NS}(y,Q^2) \\
&+\frac{\alpha_s^2(Q^2)}{4\pi R^2 Q^2}\int_{x/2}^{1/2}
  \frac{dy}{y}xP_{qq\rightarrow q}(x,y)y
  [\Sigma(y,Q^2)-q^{NS}(y,Q^2)]yq^{NS}(y,Q^2),(if~1/2\le x\le 1),
\end{aligned}
\label{full-ZRS-NS}
\end{equation}
for the non-singlet quarks,
\begin{equation}
\begin{aligned}
&Q^2\frac{dxq^s(x,Q^2)}{dQ^2} \\
&=\frac{\alpha_s(Q^2)}{2\pi}[P_{qq}\otimes q^s+P_{qg}\otimes g] \\
&-\frac{\alpha_s^2(Q^2)}{4\pi R^2 Q^2}\int_x^{1/2}
  \frac{dy}{y}xP_{gg\rightarrow q}(x,y)[ yg(y,Q^2)]^2 \\
&+\frac{\alpha_s^2(Q^2)}{4\pi R^2 Q^2}\int_{x/2}^x
  \frac{dy}{y}xP_{gg\rightarrow q}(x,y)[ yg(y,Q^2)]^2 \\
&-\frac{\alpha_s^2(Q^2)}{4\pi R^2 Q^2}\int_x^{1/2}
  \frac{dy}{y}xP_{q\overline{q}\rightarrow q}(x,y)
  y\overline{q}^s(y,Q^2)yq^s(y,Q^2) \\
&+\frac{\alpha_s^2(Q^2)}{4\pi R^2 Q^2}\int_{x/2}^{x}
  \frac{dy}{y}xP_{q\overline{q}\rightarrow q}(x,y)
  y\overline{q}^s(y,Q^2)yq^s(y,Q^2) \\
&-\frac{\alpha_s^2(Q^2)}{4\pi R^2 Q^2}\int_x^{1/2}
  \frac{dy}{y}xP_{qq\rightarrow q}(x,y)
  y[\Sigma(y,Q^2)-q^s(y,Q^2)]yq^s(y,Q^2) \\
&+\frac{\alpha_s^2(Q^2)}{4\pi R^2 Q^2}\int_{x/2}^{x}
  \frac{dy}{y}xP_{qq\rightarrow q}(x,y)
  y[\Sigma(y,Q^2)-q^s(y,Q^2)]yq^s(y,Q^2) \\
&-\frac{\alpha_s^2(Q^2)}{4\pi R^2 Q^2}\int_x^{1/2}
  \frac{dy}{y}xP_{qg\rightarrow q}(x,y)yg(y,Q^2)yq^s(y,Q^2) \\
&+\frac{\alpha_s^2(Q^2)}{4\pi R^2 Q^2}\int_{x/2}^x
  \frac{dy}{y}xP_{qg\rightarrow q}(x,y)
  yg(y,Q^2)yq^s(y,Q^2),(if~x\le 1/2), \\
&Q^2\frac{dxq^s(x,Q^2)}{dQ^2} \\
&=\frac{\alpha_s(Q^2)}{2\pi}[P_{qq}\otimes q^s+P_{qg}\otimes g] \\
&+\frac{\alpha_s^2(Q^2)}{4\pi R^2 Q^2}\int_{x/2}^{1/2}
  \frac{dy}{y}xP_{gg\rightarrow q}(x,y)[ yg(y,Q^2)]^2 \\
&+\frac{\alpha_s^2(Q^2)}{4\pi R^2 Q^2}\int_{x/2}^{1/2}
  \frac{dy}{y}xP_{q\overline{q}\rightarrow q}(x,y)
  y\overline{q}^s(y,Q^2)yq^s(y,Q^2) \\
&+\frac{\alpha_s^2(Q^2)}{4\pi R^2 Q^2}\int_{x/2}^{1/2}
  \frac{dy}{y}xP_{qq \rightarrow q}(x,y)y
  [\Sigma(y,Q^2)-q^s(y,Q^2)]yq^s(y,Q^2)\\
&+\frac{\alpha_s^2(Q^2)}{4\pi R^2 Q^2}\int_{x/2}^{1/2}
  \frac{dy}{y}xP_{qg\rightarrow q}(x,y)yg(y,Q^2)
  yq^s(y,Q^2),(if~1/2\le x\le 1),\\
\end{aligned}
\label{full-ZRS-sea}
\end{equation}
for the symmetric sea quarks and
\begin{equation}
\begin{aligned}
&Q^2\frac{dxg(x,Q^2)}{dQ^2} \\
&=\frac{\alpha_s(Q^2)}{2\pi}[P_{gq}\otimes \Sigma+P_{gg}\otimes g] \\
&-\frac{\alpha_s^2(Q^2)}{4\pi R^2 Q^2}\int_x^{1/2}
  \frac{dy}{y}xP_{gg\rightarrow g}(x,y)[ yg(y,Q^2)]^2 \\
&+\frac{\alpha_s^2(Q^2)}{4\pi R^2 Q^2}\int_{x/2}^x
  \frac{dy}{y}xP_{gg\rightarrow g}(x,y)[yg(y,Q^2)]^2 \\
&-\frac{\alpha_s^2(Q^2)}{4\pi R^2 Q^2}\int_x^{1/2}
  \frac{dy}{y}xP_{q\overline{q}\rightarrow g}(x,y)\sum_{q}[yq^s(y,Q^2)]^2 \\
&+\frac{\alpha_s^2(Q^2)}{4\pi R^2 Q^2}\int_{x/2}^x
  \frac{dy}{y}xP_{q\overline{q}\rightarrow g}(x,y)\sum_{q}[yq^s(y,Q^2)]^2 \\
&-\frac{\alpha_s^2(Q^2)}{4\pi R^2 Q^2}\int_x^{1/2}
  \frac{dy}{y}xP_{qg\rightarrow g}(x,y)y\Sigma(y,Q^2)yg(y,Q^2) \\
&+\frac{\alpha_s^2(Q^2)}{4\pi R^2 Q^2}\int_{x/2}^x
  \frac{dy}{y}xP_{qg\rightarrow g}(x,y)y\Sigma(y,Q^2)yg(y,Q^2),(if~x\le 1/2), \\
&Q^2\frac{dxg(x,Q^2)}{dQ^2} \\
&=\frac{\alpha_s(Q^2)}{2\pi}[P_{gq}\otimes \Sigma+P_{gg}\otimes g] \\
&+\frac{\alpha_s^2(Q^2)}{4\pi R^2 Q^2}\int_{x/2}^{1/2}
  \frac{dy}{y}xP_{gg\rightarrow g}(x,y)[ yg(y,Q^2)]^2 \\
&+\frac{\alpha_s^2(Q^2)}{4\pi R^2 Q^2}\int_{x/2}^{1/2}
  \frac{dy}{y}xP_{q\overline{q}\rightarrow g}(x,y)\sum_q[ yq^s(y,Q^2)]^2 \\
&+\frac{\alpha_s^2(Q^2)}{4\pi R^2 Q^2}\int_{x/2}^{1/2}
  \frac{dy}{y}xP_{qg\rightarrow g}(x,y)y\Sigma(y,Q^2)yg(y,Q^2),
  (if~1/2\le x\le1),
\end{aligned}
\label{full-ZRS-gluon}
\end{equation}
for the gluons. The linear evolution terms are from the standard DGLAP
evolution and the recombination functions $P_{ab\rightarrow c}$ are
written as
\begin{equation}
\begin{aligned}
&P_{gg\rightarrow g}(x,y)=\frac{9}{64}
  \frac{(2y-x)(72y^4-48xy^3+140x^2y^2-116x^3y+29x^4)}{xy^5}, \\
&P_{gg\rightarrow q}(x,y)=P_{gg\rightarrow \overline{q}}(x,y)
  =\frac{1}{96}\frac{(2y-x)^2(18y^2-21xy+14x^2)}{y^5}, \\
&P_{qq\rightarrow q}(x,y)=P_{\overline{q}\overline{q}\rightarrow
  \overline{q}}(x,y)=\frac{2}{9}\frac{(2y-x)^2}{y^3}, \\
&P_{q\overline{q}\rightarrow q}(x,y)=P_{q\overline{q}\rightarrow
  \overline{q}}(x,y)=\frac{1}{108}\frac{(2y-x)^2(6y^2+xy+3x^2)}{y^5}, \\
&P_{qg\rightarrow q}(x,y)=P_{\overline{q}g\rightarrow\overline{q}}(x,y)
  =\frac{1}{288}\frac{(2y-x)(140y^2-52yx+65x^2)}{y^4}, \\
&P_{qg\rightarrow g}(x,y)=P_{\overline{q}g\rightarrow g}(x,y)
  =\frac{1}{288}\frac{(2y-x)^2(304y^2-202yx+79x^2)}{xy^4}, \\
&P_{q\overline{q}\rightarrow g}(x,y)=
  \frac{4}{27}\frac{(2y-x)(18y^2-9yx+4x^2)}{xy^3}.
\end{aligned}
\label{recomb-funcs}
\end{equation}
Our numerical calculations show that the following approximation is good,
which is given by
\begin{equation}
\begin{aligned}
&P_{qq\rightarrow q}(x,y)=P_{\overline{q}\overline{q}
  \rightarrow\overline{q}}(x,y)=0, \\
&P_{q\overline{q}\rightarrow q}(x,y)=P_{q\overline{q}
  \rightarrow\overline{q}}(x,y)=0, \\
&P_{q\overline{q}\rightarrow g}(x,y)=0, \\
&P_{qg\rightarrow q}(x,y)=P_{\overline{q}g
  \rightarrow\overline{q}}(x,y)=0, \\
&P_{qg\rightarrow g}(x,y)=P_{\overline{q}g
  \rightarrow g}(x,y)=0.
\end{aligned}
\label{simpli-recomb-funcs}
\end{equation}
The reason is that the gluon density is significantly larger than the quark density at small $x$.
Figure \ref{simplification} shows the parton distribution functions
at $Q^2=5$ GeV$^2$ using the complete recombination corrections
(\ref{recomb-funcs}) and the simplified recombination corrections (\ref{simpli-recomb-funcs})
from the same initial parton distributions.
The maximum relative deviation between these two evolutions is estimated to be smaller than 5\%.

\end{document}